\documentclass[twocolumn,superscriptaddress,showpacs,showkeys,byrevtex,letterpaper,pre]{revtex4}
\usepackage{amsfonts,amsmath,amssymb}
\usepackage{epsfig}
\usepackage{graphics}
\usepackage{dcolumn}
\usepackage{bm}
\begin{document}
\title{Multifractal Analysis of Polyalanines Time Series}

\author{P.\, H.\, Figueir\^edo}
\email{phugof@df.ufrpe.br}
 \affiliation{Departamento de F\'{\i}sica, Universidade Federal Rural de
Pernambuco, 52171-900, Recife, Pernambuco, Brazil}

\author{E.\ Nogueira Jr.}
\email{enogue@fis.ufba.br}
\affiliation{Instituto de F\'{\i}sica, Universidade Federal da Bahia
40130-240, Salvador, Bahia, Brazil.}

\author{M.\, A.\, Moret}
\email{moret@cairu.br}
\affiliation{Departamento de F\'{\i}sica, Universidade Estadual de Feira de Santana, 44031-460, Feira de
Santana, Bahia, Brazil.}

\author{S\'ergio Coutinho}
\email{sergio@ufpe.br}
\affiliation{Laborat\'orio de F\'{i}sica Te\'{o}rica e Computacional, Departamento de
F\'{i}sica, Universidade Federal de Pernambuco, 50670-901, Recife, Pernambuco, Brazil}

\date{\today}

\begin{abstract}
Multifractal properties of the energy time series of short $\alpha$-helix structures, specifically from a polyalanine family, are investigated through the MF-DFA technique ({\it{multifractal detrended fluctuation analysis}}). Estimates for the generalized Hurst exponent $h(q)$ and its associated multifractal exponents $\tau(q)$ are obtained for several series generated by numerical simulations of molecular dynamics in different systems from distinct initial conformations. All simulations were performed using the GROMOS force field, implemented in the program THOR. The main results have shown that all series exhibit multifractal behavior depending on the number of residues and temperature. Moreover, the multifractal spectra reveal important aspects on the time evolution of the system and suggest that the nucleation process of the secondary structures during the visits on the energy hyper-surface is an essential feature of the folding process.
\end{abstract}


\maketitle
\section{Introduction}

Over the past few years, the statistical analysis of self-similar time series has become established as an important tool for investigating several natural phenomena. In general, a large part of these studies has been devoted to characterizing the complex statistical fluctuations shown by these series. Such fluctuations are associated to long-range correlations among the dynamic variables present in these series, and which obey the behavior usually described by fractal power-law decay \cite{Stanley99}.

One of the difficulties encountered in these investigations is related to the fact that the series may contain heterogeneous properties imposing certain statistical trends over itself. In other words, these series are not stationary \cite{Chianca2005}. Therefore, it is necessary to employ a technique capable of accounting, for this, since these trends may influence the correlations that exist in the series.

Two techniques have proved successful in eliminating these trends in time series: the wavelet transform modulus maxima (WTMM) \cite{Arneodo1995,Manimaran2005} and the detrended fluctuation analysis (DFA) \cite{Peng1994}. Both techniques are based on local polynomial regression in order to eliminate local trends present in different segments of the series. The DFA technique has been particularly efficient for a large range of areas such as: DNA sequences \cite{Peng1992}; heartbeat analysis \cite{Ivanov1999}; economy \cite{Rogerio2003}; seismology \cite{Telesca2001,Telesca2004}; meteorology \cite{Govindan2003}; astrophysics \cite{Zebende2003}, among others.

Basically, the option of applying the DFA technique to these studies stems from its easy implementation. Moreover, it is a tool that allows the role of trends in stationary time series to be analyzed, as well as efficiently estimating the long-range correlations through a single parameter: the scale exponent $\alpha$. Is important to emphasize that the type of correlation present in the stationary series depends on the value found for exponent $\alpha$. In this way, for $\alpha=0.5$ the signal is uncorrelated (white noise or Gaussian), while for $\alpha < 0.5$ there is anti-correlation (anti-persistence) and for $\alpha > 0.5$ there is correlation (persistence) \cite{Vicsek}.

Several attempts to apply the DFA technique to non-stationary time series (series affected by local trends) have not provided satisfactory results. Fundamentally, this has occurred because these series are not entirely characterized by a unique scale exponent $\alpha$, since different segments possess fluctuations characterized by distinct values of $\alpha$. In this case, the correct formalism for obtaining the distribution of scale exponents is multifractal analysis \cite{Vicsek}.

Recently, the number of works focusing on the characteristics of the multifractal aspects of non-stationary time series has grown, particularly those based on experimental data \cite{Pawel2006}. Outstanding among these are the applications of the generalized DFA technique, known as {\it Multifractal Detrended Fluctuation Analysis} (MF-DFA) as proposed by Kanterlhardt and collaborators \cite{Kantelhardt2002}, for a wide range of applications such as: DNA sequences \cite{Nogueira2002}, meteorology \cite{Kavasseri}, seismology \cite{Telesca} and others. It should also be remembered that two factors influence the use of the MF-DFA technique: its effortless implementation, and its excellent performance in obtaining results, related to both artificial and real data, when compared to the performance of the Wavelet Transform process, for the same systems \cite{Pawel2006}.

One highly relevant problem in molecular biology within this context, is linked to studies concerning the protein folding process through the characterization of its potential energy landscape. Such landscapes constitute a satisfactory representation of the potential energy for interaction among the various system's microscopic freedom degrees \cite{Rainer2007}.

In general, the adopted strategies are based on the assumption that the energy landscapes of proteins are complex, since being time dependent, they present a rugose structure, with many maxima and minima separated by barriers of varying heights. These properties imply complex evolutionary dynamics, in which the system experiences a variety of time scales \cite{PRE2008Mazzoni,Lorenzo 2006}.

Previous studies, using molecular dynamics simulations (MOIL program) and a variational method of fractal analysis to study the fractal properties of time series of the potential energy of molecular systems such as myoglobin; polyalanines, among others, were conducted by Lidar and collaborators \cite{Lidar1999}. Basically, they investigated systems that were subjected to a temperature $ T = $ 300K and a simulation time in the range $ 10 <t <$ 25 ps.

Their results suggest that the value of the fractal dimension (the rugosity exponent) slightly depends on temperature and the presence of $\alpha$-helix structures smoothes the rugosity of the series. Furthermore, there was evidence of universal behavior, i.e. the rugosity of different systems is described by the same fractal dimension. Recently, Hegger et al.  \cite{Rainer2007} analyzed time series extracted from molecular dynamics simulations (GROMACS program) at a temeperature of $T=300$K, for polyalanines with the number $N$ of amino acids ranging between $3$ and $10$, reaching simulation time of $100$ ns.

Considering that this time series represent the dynamic trajectories followed by the system, these authors found that the effective fractal-dimension of such trajectories decreases with the chain size. According to them, such behavior occurs due to a stabilizing effect of the hydrogen bonds on the protein secondary structure ($\alpha$-helix) smoothing rugosities on the trajectories. Confirming whether this scenario is able to survive careful fractal analysis, searching for fine details of the time series fluctuations, has become a central problem to be clarified.

The present work introduces an approach, which combines molecular dynamics simulations with MF-DFA to characterize the rugosity of potential energy profiles, for polyalanine molecules. By considering these profiles as energy time series, we investigate the effects produced on the trajectories traced over the hyper-surface of the potential energy, when the size and temperature of the system is changed. In particular, we will show that the manner in which the system visits the phase-space in its dynamic evolution significantly depends on both temperature equilibrium and the nucleation of secondary structures in polyalanines.

This article is organized as follows: Section II presents the molecular dynamics simulations, the energy time series, and the energy dependence on temperature $T$ and number of amino acids $N$. Section III presents the multifractal spectra, obtained from the MF-DFA technique, associated to each different polyalanine time series. The effects caused by changes on the size of the chain and temperature and the presence of secondary structures on the spectra are discussed. Finally, Section IV presents our
conclusions.

\section{Molecular Dynamics and Time Series}

Molecular dynamics simulations have been extensively used to study the problem of protein folding \cite{Moret2005}. In general, these simulations involve considerable computational effort, since the integration of the equations of motion must be made for a system with many particles.

In the case of molecular systems, it is known that such structures can take on a great number of configurations, which grow with the number of degrees of freedom of the system. Therefore, molecular dynamics calculations for protein systems necessarily require the definition of effective potentials, from which the resulting force that acts on each particle is determined.

In this work, the numerical molecular dynamics calculations were performed with the aid of an efficient computer code: the THOR program \cite{Pascutti99}, developed to investigate structures of biological interest, such as proteins. The code includes the GROMOS force field \cite{Gunsteren1987} in its architecture, used to simulate the atomic interactions in the molecule.

In the THOR program, the conformational energy of the molecule is made up of a sum of bonded and nonbonded terms. In such approach, only hydrogen atoms covalently bonded to oxygen or to nitrogen are considered explicitly, whereas CH1, CH2, and CH3 groups are assumed to be an atomic unit. Thus, we analyze the changes of the following energy function:
\begin{equation}
    E=E_H+E_\theta+E_\phi+E_\varphi+E_{LJ}+E_{el}
\end{equation}
or explicitly,
\begin{eqnarray}
E &=&\frac{1}{2}\sum_{k}K_{b_{k}}(r_{k}+r_{0})^{2}+ \frac{1}{2}\sum_{l}K_{\theta _{l}}(\theta _{l}+\theta
_{0})^{2}+ \nonumber \\ &&+\frac{1}{2}\sum_{m}K_{{\phi}_{m}}(\phi_{m}+\phi_{0})^{2}+ \nonumber \\
&&+\sum_{n}K_{\varphi_{n}}(1+\cos (m\varphi _{n}+\varphi _{0}))+\\ &&+\sum_{i<j}\left[
\frac{C_{12}(i,j)}{r_{i.j}^{12}}-\frac{C_{6}(i,j)}{r_{i,j}^{6}}\right] +\frac{1}{4\pi \varepsilon
_{0}\varepsilon _{r}} \sum_{i<j}\frac{q_{i}q_{j}}{r_{i,j}}.  \nonumber
\end{eqnarray}
where $E_H$ is the Hook potential, $E_\theta$ is the angular potential, $E_\phi$ is the umproper potential, $E_\varphi$ is the dihedral-angle potential, $E_{LJ}$ is the Lennard-Jones potential, and $E_{el}$ is the Coulomb potential term (see definitions and used parameters in \cite{Gunsteren1987,Pascutti99}).

Specifically, we simulate polyalanine structures with a different number of residues at different equilibrium temperatures and initial conformations. Polyalanines are used as prototypes to study the folding process of structures in $\alpha$-helix conformations. In this dynamic, the electric dipole moments arising from the electric unbalance between the peptide bond of $NH$ and $CO$ groups, the hydrogen bridges bonds and the van der Waals interactions, are key ingredients in the cooperative effect responsible for forming such structures, and which becomes accelerated with the increasing number of amino acids in the protein.

Thus, as pointed out by Shoemaker and collaborators \cite{Shoemaker1987}, Moret and collaborators \cite{Moret2002} and Rogers \cite{Rogers1989}, a critical minimum number of amino acids is necessary so that these configurations may be observed. Furthermore, there is an upper critical number due to destabilization brought on by entropic effects.

For the numerical calculations, a similar protocol was adopted in all cases. The initial temperature started at $T_i=1$K, heating the system at a rate of $5$K per step (ps) in order to reach the desired equilibrium temperature. Three equilibrium temperatures were considered: $T=275$K, $T=300$K and $T=325$K, in a continuous medium, described by a relative dielectric constant $\epsilon_r=2$. The increases in the time dynamic was $5\, 10^{-4}$ ps, and for all simulations $N_{\textsf{step}}=5\,10^8$ steps were performed, to achieve a time of the order of $25$ns. In calculating the time, the interval associated with the thermalization of the system was discarded.
\begin{figure}
\begin{center}
\includegraphics*[width=\linewidth]{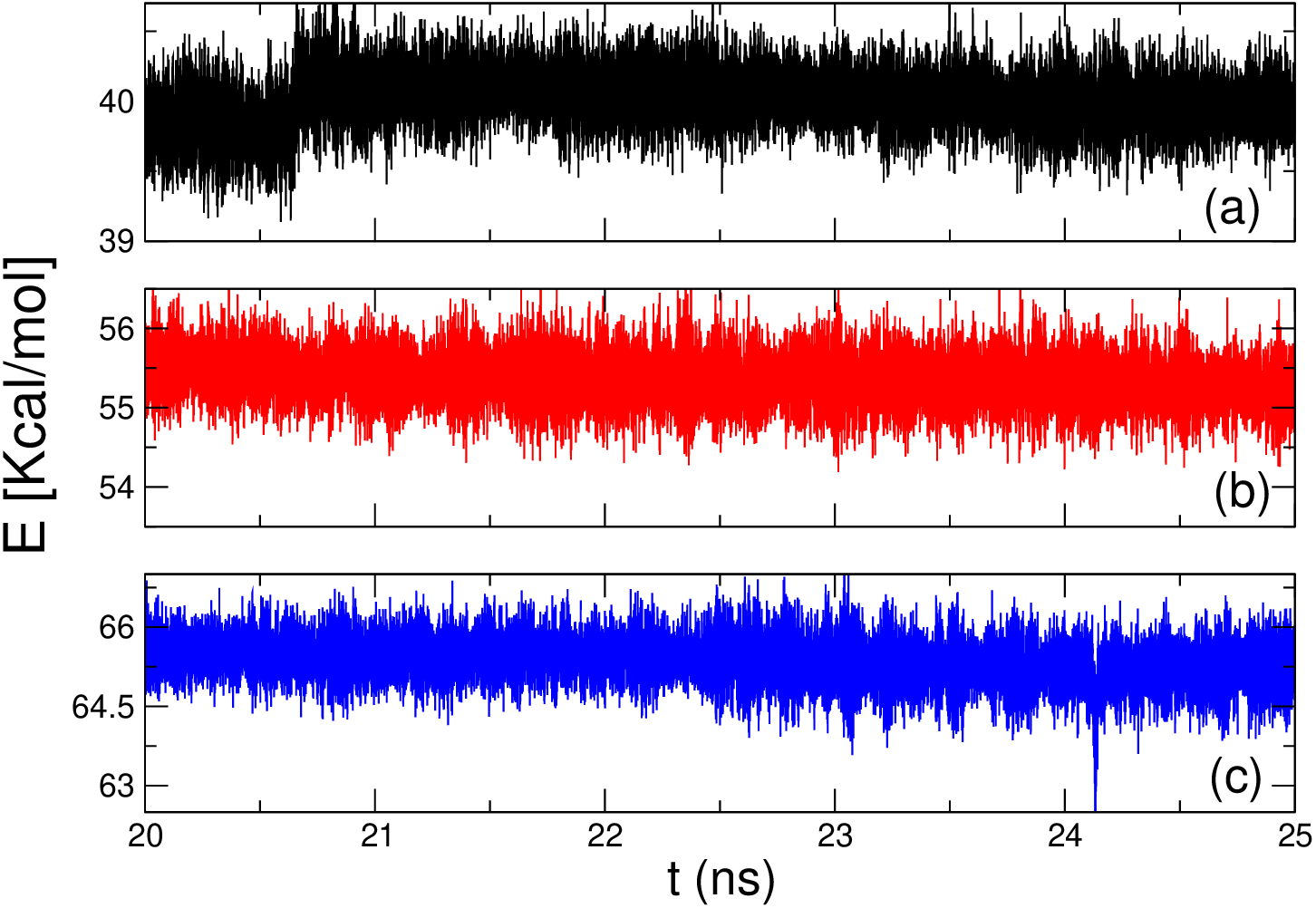}
\end{center}
\caption{ Potential energy time-series of polyalanines with different numbers of residues (a) N=10 (black), (b)  N = 15 (red), and  (c) N = 18 (blue). In all cases the thermal equilibrium temperature is $T = 300$K.
\label{fig:figura1}}
\end{figure}

Figure (\ref{fig:figura1}) shows in the last $5$ns of observation for the potential energy time series associated to polyalanine structures in $T = 300$K and $N=10$, $15$, and $18$. It was noted that in all cases examined, the series showed the typical rugosity observed in other complex phenomena described by self-affine time series \cite{Vicsek}. It should be emphasized that for all temperature values, calculations were reached with values of $N$ between 8 and 18 residues, the results of which display similar behavior to those presented in Figure (\ref{fig:figura1}).

\section{MF-DFA -- Multifractal Spectra}

Once the time series of polyalanines have been obtained and the presence of rugosity has been observed, a careful characterization of the statistical fluctuations embedded in the series should be performed in order to obtain information on the dynamic behavior of the system. In this work, the MF-DFA method is applied, along the following steps \cite{Kantelhardt2002}:

\begin{enumerate}
\item Consider a time series $u(i)$, $i=\{1...N_{max}\}$, over a compact support, and determine its profile (integrated set), i.e.,
\begin{eqnarray}
y(i) = {\sum\limits_{k = 1}^{i} {[u(k)-<u>]}}, \label{eq:serie-integracao}
\end{eqnarray}
where $<u>$ is the mean taken over the original series $u(i)$;
\item Divide the profile in $N_s$ disjointed segments of equal size $s$, calculate the local trend through a polynomial adjustment of order $m$, represented by the variable $y_{\nu}(i)$, at each segment. Since the length $N$ of the series is often not a multiple of the considered time scale $s$, a short part at the end of the profile may remain. In order not to disregard this part of the series, the same procedure is repeated starting from the opposite end. Thereby, $2N_s$ segments are obtained altogether;
\item Determine the fluctuation variance,
\begin{equation}\label{eq:serieF2}
    F^2(s,\nu) \equiv \frac{1}{s} {{\sum\limits_{i = 1}^s
\{{{{ y[(\nu-1)s +i]-y_{\nu}(i)}} }\}^2}},
\end{equation}
with $\nu =\{1,...,N_s\}$, associated to each segment. Notice that in this step, polynomial trends of the order $m$ were eliminated from the profile.
\item Calculate the mean values of all segments, to obtain the fluctuation function of the order $q$:
\begin{eqnarray}
F_{q}(s) \equiv \left\{\frac{1}{2N_s}\sum\limits_{\nu = 1}^{2N_s}
[F^2(s,\nu)]^{q/2}\right\}^{1/q},\label{eq:serieFq}
\end{eqnarray}
where, in general, the variable $q$ assumes real values, except zero.
\end{enumerate}

The characteristic property of function $F_{q}(s)$ is its scale behavior, i.e. if the time series $u(i)$ possess long-range correlations, then for increasing values of $s$, function $F_{q}(s)$ also grows, following a power law of the type:
\begin{eqnarray}
        F_q(s) \sim s^{h(q)}.
\label{eq:serieFq-tau}
\end{eqnarray}
Therefore, the main result obtained with the MF-DFA method is a family of exponents $h(q)$, called the generalized Hurst exponents. For a genuine multifractal series these exponents form a decreasing function of $q$, if on the other hand, the signal is monofractal $h(q)$ = constant. Moreover, if $q<0$, $h(q)$ captures the properties of small fluctuations, then for $q>0$ large fluctuations are dominant. Particularly, when $q=2$, $h(2)=H$ is the classical Hurst exponent.

Finally, the multifractal spectrum of measures can be obtained through a simple relation between the exponent $h(q)$ and the multifractal scale exponent $\tau(q)$, defined by multifractal formalism \cite{Kantelhardt2002}:
\begin{eqnarray}
    \tau(q) \equiv qh(q) - 1.
\label{eq:tauq}
\end{eqnarray}
The function $\tau(q)$ is one of the most used representations of multifractal spectra, related to time series.

\begin{figure}[thbp]
\begin{center}
\includegraphics*[width=\linewidth]{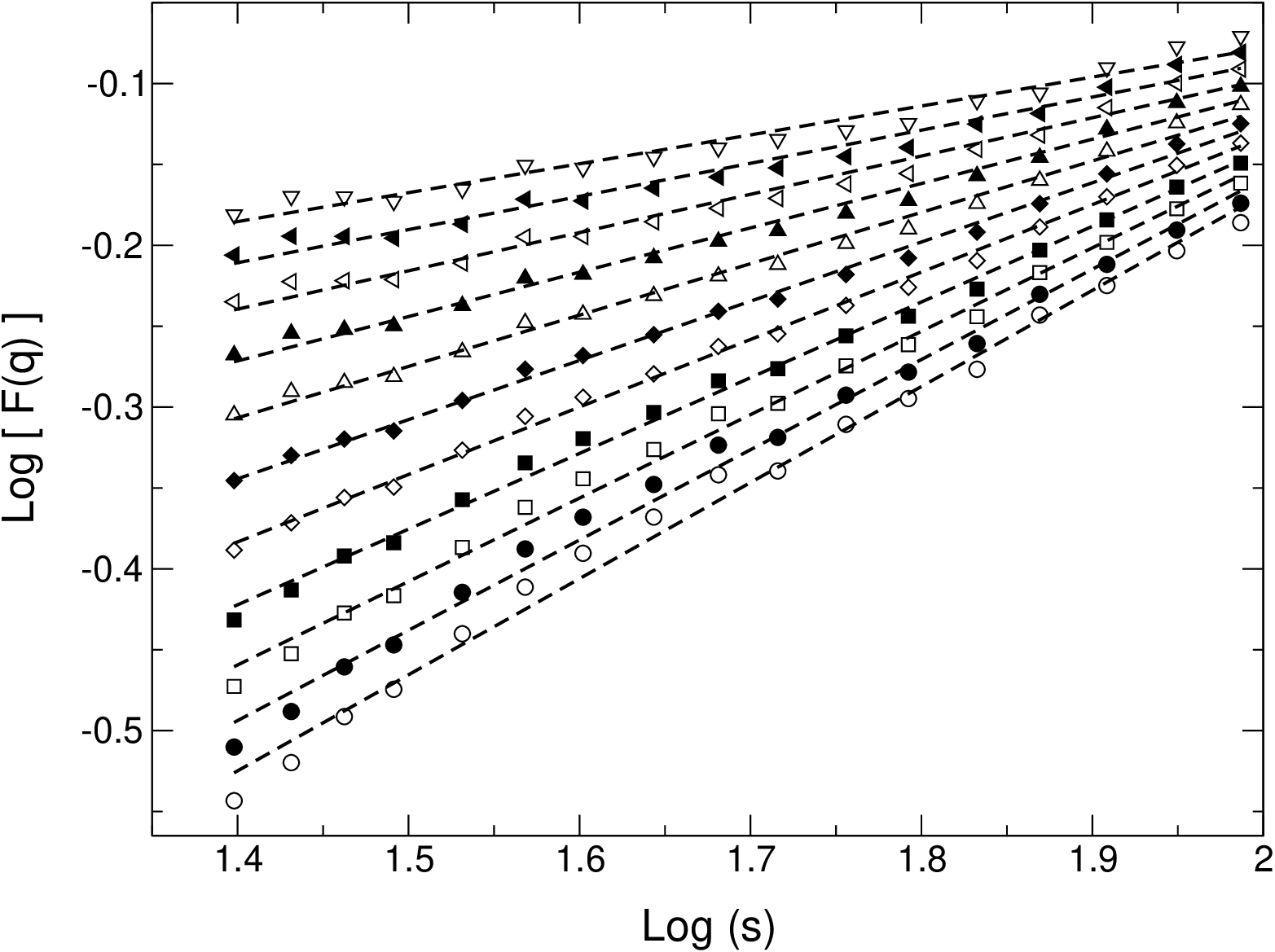}
\end{center}
\caption{Logarithmic of the fluctuation function $F_q(s)$ against $\log s$ with the $q$ parameter
$-5<q<5$ (step 1 from top to bottom) for polyalanine energy time series with $N=18$ residues, and thermal equilibrium temperature of $T=300$K.}\label{fig:figura2}
\end{figure}

Furthermore, typical results are presented obtained using the MF-DFA technique to investigate the different time series of the potential energy of polyalanines as described in Section II. Figure (\ref{fig:figura2}) represents the behavior of the logarithmic of the fluctuation function $\log F_q(s)$ as a function of $\log s$ and the parameter $q$, for the series with $N=18$ residues shown in Figure (\ref{fig:figura1}). The scale values were chosen in the range $20<s<100$ and the trends were approximated by a polynomial of order $m=4$. As can be observed, the estimates obtained for the linear adjustment of the data satisfactorily meet the behavior of the scale provided by Equation (\ref{eq:serieFq-tau}).

\begin{figure}[thbp]
\begin{center}
\includegraphics*[width=\linewidth]{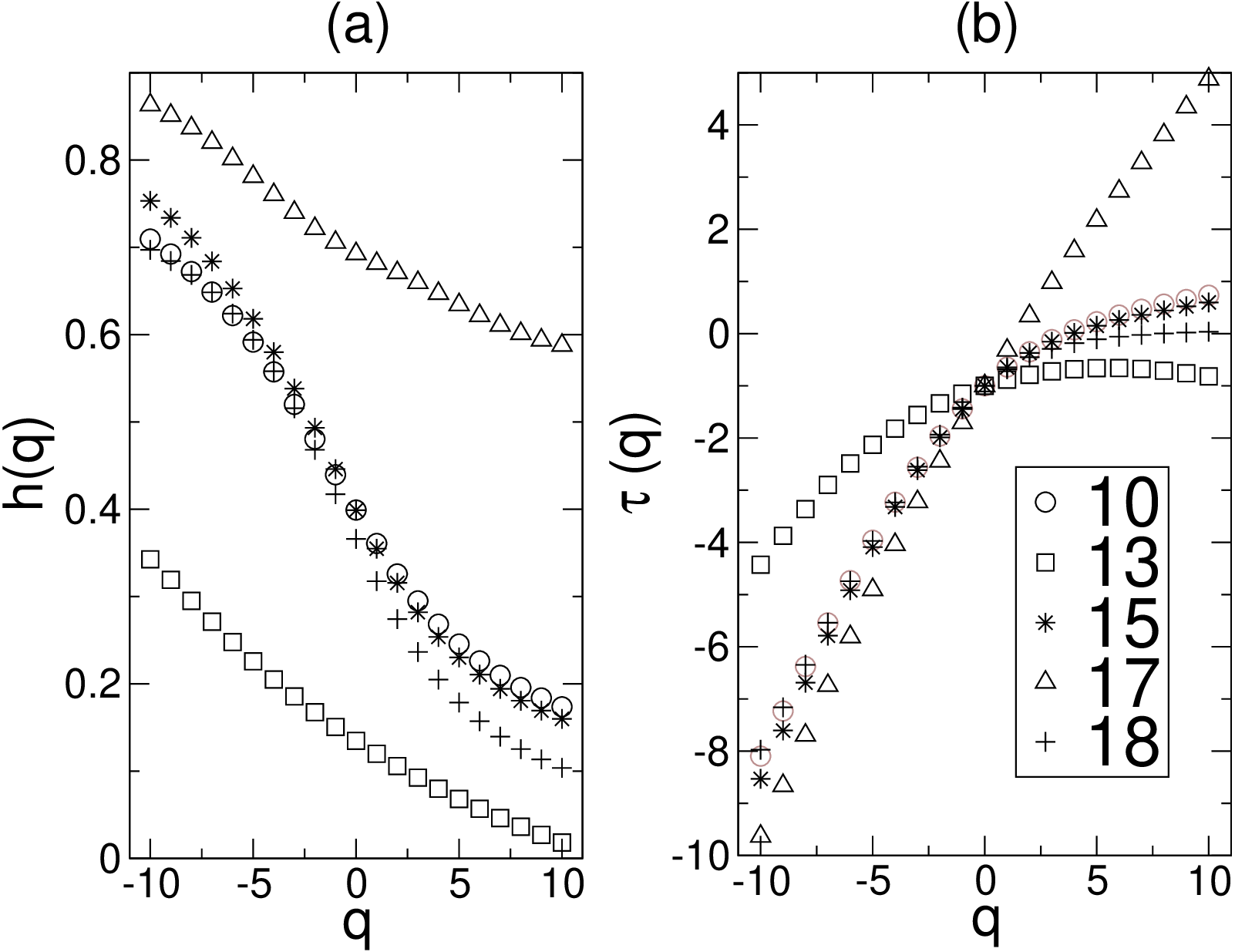}
\end{center}
\caption{(a) Generalized Hurst exponents $h(q)$ dependence on the parameter $q$ and (b) multifractal spectrum $\tau (q)$ dependence on $q$ for polyalanines with a different number of residues, thermal equilibrium temperature of $T=300$K.}\label{fig:figura3}
\end{figure}

Figure (\ref{fig:figura3}a) presents the corresponding exponents $h(q)$ (values of the slopes of the straight lines fitted to the data) as a function of $q$, while Figure (\ref{fig:figura3}b)  presents the associated multifractal spectrum $\tau(q)$. In general terms, it may be stated that the results indicate that the time series investigated exhibit typical multifractal behavior ($\tau(q)$ is not a linear function of $ q $), which depends on the number of residues $N$ and the thermal temperature $T$ of system.

In Figure (\ref{fig:figura3}) different regimes of correlation may be observed: for $N=17$, the series is completely correlated; while for $N=10$, $15$ and $18$, there is a mixed system, i.e. a strong anti-correlation when $q>0$ and correlation, for some values of $q<0$. In particular, when $N=13$ the series is totally anti-correlated. According with reference \cite{Moret2002} $N=13$ is the  critical number of residues associated with the formation of $\alpha$-helix in $T=300$K.

Since for this value of $N$, the potential energy is a global minimum then we may consider that the nucleation of secondary structures alters the dynamics of the system for an anti-correlated regime, thus overcoming the growth trend of the energy induced by thermal agitation and the increase of residues. In addition, as in Figure (\ref{fig:figura3}b), all spectra $\tau(q)$ exhibit typical
multifractal properties.

\section{Conclusions}

In this work, we have studied the multifractal properties of time series of the potential energy of polyalanines. Protein chains were analyzed with different numbers of residues at three equilibrium temperatures. The research was conducted using an approach that combines molecular dynamics with MF-DFA, a technique of statistical analysis, which enabled us to characterize the rugosity associated with the temporal correlation among the dynamic variables of the series.

Our results corroborate some of those obtained by Hegger \emph{et.al} \cite{Rainer2007} and Lidar and collaborators \cite{Lidar1999}, such as the influence of time and the presence of the $\alpha$-helix in the rugosity of the time series. However, they also indicate that the other findings have not been confirmed, since the time simulation they used is much shorter than that used in this study and so insufficient to observe the formation of secondary structures. Also, the fractal analysis technique employed by these authors, which does not deal adequately with the existence of trends in the series, has not allowed to capture the subtler details of the spectra.

Indeed, the results obtained in this study indicate that all the series examined exhibit typical multifractal behavior, which depends both on the number of residues $N$, and the temperature $T$ of the system, and that these multifractal properties, represented by $\tau(q)$ spectra or, similarly, the generalized Hurst exponents $h(q)$, reveal important aspects of the temporal evolution of the system.

It was found, for example, that when the number of residues approaches the critical number of residues $N_c$, associated with the formation of a significant amount of secondary structures, the temporal correlation regime of the system is changed. In the case $N_c=13$ and $T=300$K, the system is totally anti-correlated, the spectrum $\tau(q)$ is truly multifractal and rugosity is more pronounced in the region of small fluctuations ($q<0$), as seen in Figure (\ref{fig:figura3}a). For other values of  $N$, the results confirm that the two regimes of correlation are present in the series.

Recently, Moret and collaborators \cite{Moret2001a} conducted an analysis of the spectra (profiles) of the potential energy of proteins, in function of the number of dihedral angles  $\phi$ and $\psi$, and found that these profiles are described by a real multifractal $f(\alpha)$ spectra. They also found that the $f(\alpha)$ spectra were sensitive to the number of degrees of freedom of the system, thus illustrating that the dimensionality of the phase space influences the accessibility of parts of the hyper-surface of the potential energy, since the proteins adopt conformations in the phase space only in the permitted regions of the spectrum $f(\alpha)$.

This behavior allows an alternative explanation for the dynamics of the clew of a protein, because it suggests the existence of preferential folding trajectories along the energy hyper-surface, i.e. in the search for its native state, proteins need not visit all the accessible states in the space phase, but only those associated with the spectrum $f(\alpha)$.

The MF-DFA method applied to the time series of the potential energy of polyalanines, has enabled this study to reveal important aspects concerning the wealth and complexity associated with the temporal evolution of these systems, in the search for its native state.

In fact, according to the number of residues and the temperature, it was shown that the trajectory of the protein, to visit its phase space dynamically, is guided mainly by the influence of secondary structures, which are formed over the time simulation, probing the hyper-surface of the conformational energy at different time scales. As a result, the energy time series exhibit multifractal long-range correlations.

Therefore, our results support an alternative explanation of the so-called Levinthal paradox
\cite{Levinthal}, because in this scenario, the protein in its dynamic evolution, is being influenced by the emergence of intermediate structures, which gradually, by successive increases in conformational stability, bypass the trajectories by way of preferential folding. Consequently, the extreme ease with which a protein is folded, despite the huge number of possible configurations, may be attributed to a succession of events, which it experiments, on a multifractal space-time energy hyper-surface.

\begin{acknowledgments}
This work was partially supported by the Brazilian federal grant agencies CNPq and CAPES, and from FACEPE (Pernambuco state grant agency) under the grants PRONEX EDT 0012-05.03/04 and APQ-0203-1.05/08).
\end{acknowledgments}

\end{document}